\documentclass[floatfix,secnumarabic,amssymb,nobibnotes,nofootinbib,preprint,aps,pra,showpacs]{revtex4-1}
\usepackage{epsfig}
\usepackage{amssymb}
\usepackage[colorlinks=true, citecolor=blue, linkcolor=blue, urlcolor=black]{hyperref}
\usepackage{cleveref}
\usepackage{times}
\usepackage{amssymb}
\usepackage{amsmath}
\usepackage{mathrsfs}
\usepackage{graphicx}
\usepackage{epsfig}
\usepackage{dcolumn}
\usepackage{color}
\usepackage{bm}
\usepackage{graphics,psfrag}
\usepackage{graphicx,psfrag}
\usepackage{braket}
\newcommand{\be}{\begin{equation}}
\newcommand{\ee}{\end{equation}}
\newcommand{\bea}{\begin{eqnarray}}
\newcommand{\eea}{\end{eqnarray}}
\newcommand{\ba}[1]{\begin{array}{#1}}
\newcommand{\ea}{\end{array}}

\usepackage{braket}
\usepackage{soul}

\begin{document}

\title{Realizing negative index of refraction in an ensemble of ground-state polar molecules with lasers}
 \author{Dibyendu Sardar$^{1}$, Sauvik Roy$^{2}$, Ghanasyam Remesh$^{3}$, Subhasish Dutta Gupta $^{3,4}$ and  Bimalendu Deb $^{1}$\footnote{Corresponding author, e-mail: msbd@iacs.res.in}}
 \affiliation{$^{1}$School of Physical Sciences, Indian Association for the Cultivation of Science (IACS), Jadavpur, Kolkata 700032, INDIA.
 \\$^{2}$Indian Institute of Science Education and Research Kolkata, Mohanpur-741 246, India.
 \\$^{3}$School of Physics, University of Hyderabad, Hyderabad-500046, India.
 \\$^{4}$Tata Centre for Interdisciplinary Sciences, TIFRH, Hyderabad 500107, India.
}
\def\zbf#1{{\bf {#1}}}
\def\bfm#1{\mbox{\boldmath $#1$}}
\def\hf{\frac{1}{2}}

\begin{abstract}
We propose a coherent optical method for creating negative refractive index (NRI) for  a gaseous ensemble of ground-state  polar molecules possessing both permanent electric and magnetic moments. Exploiting the pure rotational transition between the two lowest rotational levels of the ground vibrational state one can generate two dressed states of mixed parity using a microwave laser. These dressed states are then used as the two lower states of a $\Lambda$-type three-level scheme using two infrared lasers to couple them to another ro-vibrational level in the ground-state manifold. One of the infrared lasers is used as a weak probe and the other as a control field with a fixed detuning. We take into account local-field effects on the dielectric response of the medium at the probe frequency in terms of Clausius-Mossoti relation. We extract magnetic response out of purely dielectric response and calculate the magnetic permeability of the medium in terms of dielectric susceptibility. Our results show that 
there is a small frequency window near the two-photon resonance where both electric permittivity and magnetic permeability are negative with vanishingly small absorption. The figure of merit for the  medium is shown to exceed unity. 
We interpret our results in terms of the proximity of EIT and  quantum interference. We discuss the possible realization of our method using  cold polar molecules that are recently experimentally produced. 
\end{abstract}

\maketitle
\section{Introduction}

Half-a-century ago,  Russian Physicist Victor G. Veselago \cite{vaselago} theoretically predicted the possibility of an exotic material possessing simultaneously both negative electric permittivity $\epsilon_r$ and magnetic permeability $\mu_r$. Based on fundamental principles of classical electrodynamics, he showed that such materials must have negative index of refraction. Materials exhibiting negative index are not found in nature. Nearly three decades after Veselago's seminal work, Sir John Pendry  first showed a practical way to design such a negative index material (NIM)  using split ring resonators (SRR) and metal wires \cite{pendry:1999}.
 Research on NRI got a tremendous boost when in 2000 Sir Pendry showed that a slab of NRI material can amplify evanescent waves and thereby can lead to perfect lensing \cite{pendry}. Note that focusing of propagating waves by such a slab had been shown by Veselago \cite{vaselago}. NRI was first experimentally demonstrated in 2001 by Smith {\it et al.} \cite{smith} in microwave frequency regime using an artificially created periodic array of SRRs and metal wires. Since then,  there has been a flurry of research activities in negative index materials and their optical properties \cite{padilla2006}. These designer materials are also known as left-handed material (LHM) or metamaterial, as the electric and magnetic field vectors and the propagation vector form a left-handed triplet \cite{vaselago, smith, sdgbook}. These materials exhibit surprising and counterintuitive electromagnetic and optical effects like propagation of phase opposite to energy flow, reversal of both Doppler shift and Cherenkov radiation \cite{vaselago}, amplification of evanescent waves \cite{pendry,feise}, negative Goos-H\"{a}nchen shift \cite{berman}, anomalous refraction \cite{vaselago}, sub-wavelength focusing and superresolution \cite{pendry,chen,fang}, reversed circular Bragg phenomena \cite{lakhtakia}, photon helicity inversion \cite{shen1},  $H$-field circulation and inverted $E$-field line in propagating structures \cite{krowne}, unusual photon tunneling effects \cite{zhang}. Several other effects like switched field intensity locations in anisotropic transmission structures \cite{krowne1}, fast and slow light \cite{sdgprb,manga2004}, lasing spasers \cite{zheludev} and optical nanocircuits \cite{Engheta} were also reported. There have been several other ways to realize left-handed materials based on classical electromagnetic principles, including chiral material \cite{pendry1,yannopapas,tretyakov,mackay} and photonic crystal structures \cite{berrier,thommen,he}. 

\par
 Most of the metamaterials \cite{padilla2006,ramakrishna_2005} created  so far using solid-state elements exhibit negative index in the long wavelength domain, namely microwave and tera-Hertz \cite{zhangtz,padilla2006} regime. Moreover, because of the underlying lattice structures, most of these materials are anisotropic in nature. Currently, there is tremendous research interest in designing new isotropic metamaterials at short wavelength or optical regime. In recent years coherent manipulation of the absorptive and dispersive properties of a medium of atomic or molecular gases are being theoretically explored as a route to achieve isotropic negative index at optical frequency domain.  In this context, Oktel and M\"{u}stecaplio\v{g}lu \cite{oktel} first proposed a three-level $\Lambda$-type scheme based on electromagnetically induced transparency (EIT) in order to achieve NRI. Since then, several coherent optical schemes using multilevel atoms or molecules have been proposed
\cite{shen1,kastel,kastel1,sikes,orth, zhang1, zubairy, othman,krowne3,budriga}.  K\"{a}stel \cite{kastel} and Thommen \cite{thommen1} have  suggested that a LHM can be obtained in an atomic four level system. Even a five-level atomic system  has been studied \cite{othman} for this purpose.  In recent years, there have been several suggestions \cite{zhao,sczhao,shen2,dutta} for creating LHM  with atomic gases, using  EIT and gain mechanism.   Despite all these theoretical works, negative index in an atomic or molecular system  is yet to be realized experimentally.

\par
\begin{figure}
\center
\includegraphics[width=0.8\linewidth]{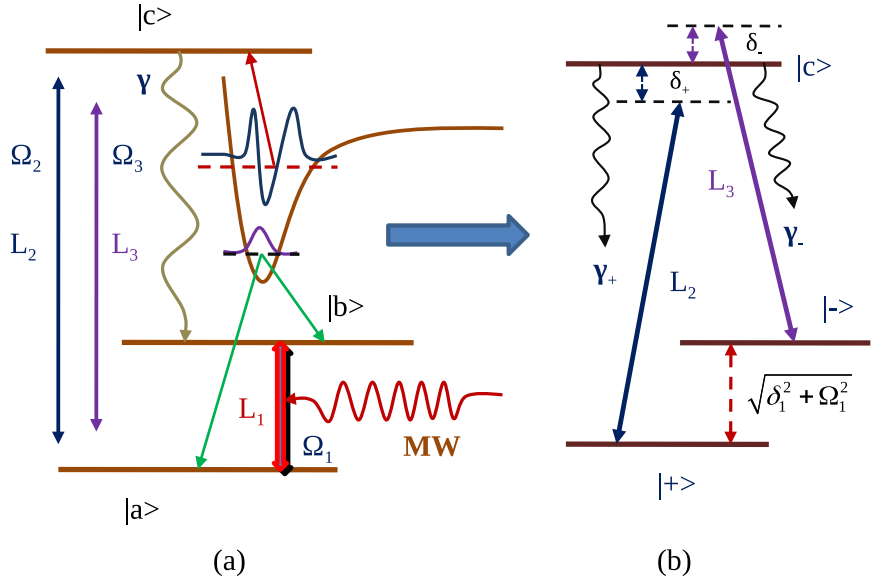}
\caption{ Schematic diagram showing a three-level polar molecular system in electronic ground-state potential.  $\ket a$, $\ket b$ and $\ket c$ represent three bare ro-vibrational levels.  A strong microwave (MW) laser L$_1$  coupling $\ket a$ and $\ket b$ with Rabi coupling $\Omega_1$ produces two dressed states $\ket +$ and $\ket-$ of mixed parity, with $\ket -$ being chosen to remain always higher in energy than $\ket+$ (see text). The two infrared lasers L$_2$ and L$_3$ corresponding to Rabi couplings $\Omega_2$ and $\Omega_3$   are off-resonant to the transition $\ket b\leftrightarrow \ket c$ in bare-state picture (a), but they  come to near-resonance with the transitions $\ket +\leftrightarrow \ket c$   
and $\ket -\leftrightarrow \ket c $, respectively, forming a $\Lambda$-type three-level system in dressed-state picture (b). $\delta_+$ and  $\delta_-$ are the detuning of L$_2$ and L$_3$ from the respective  transitions. The level $\ket c $ can radiatively decay to the level $\ket b$ with decay constant $\gamma$. $\gamma_{+}$ and $\gamma_{-}$ are the decay constants for transitions $\ket c \rightarrow \ket +$ and 
$\ket c \rightarrow \ket -$, respectively. In our model calculations (see text), we assume that $\ket{a}$ and $\ket{b}$ are in the ground vibrational level while $\ket{c}$ has a different vibrational level as shown in a representative molecular potential plot in the subplot (a).}
\label{potential}
\end{figure}

 With the possibility of achieving negative refraction with suppressed absorption near EIT in mind, here we propose a new approach using a  polar molecular gas possessing both permanent magnetic and electric dipole moments. Our system is schematically shown in Fig.\ref{potential}. 
We consider  microwave-dressed states $\ket+$ and $\ket -$ which are linear superposition of two bare states $\ket a$ and $\ket b$ having opposite parity. So, the dressed states have mixed parity and both electric and magnetic dipole allowed transitions may take place between either of the two  dressed states and a third level $\ket c$ with a definite parity. Our scheme allows us to manipulate the values of $\epsilon_r$ and $\mu_r$ over a certain frequency band in infrared regime while minimizing the absorption suitably. Considering one of the infrared lasers (say, L$_3$)  as the control field tuned between the upper dressed state $\ket -$ and $\ket c$ while the other laser (L$_2$) treated as a probe is tuned between the lower dressed state $\ket +$ and $\ket c $. Keeping the detuning of  the control field fixed, we scan the frequency of the probe across the two-photon resonance. Our results show that, close to two-photon resonance there is a small probe detuning regime where both the real parts of electric 
permittivity and magnetic 
permeability are negative with small absorption, and so the real part of refractive index $n$ becomes negative in this domain. The figure of merit, defined as the absolute magnitude of the ratio $ n_r/n_i$ of the real part $n_r$ to the imaginary part $n_i$ of $n$, can be as large as 10. The vanishingly small absorption in 
this domain can be attributed to quantum interference between the two lower states $\ket+$ and $\ket -$ near the electromagnetically induced transparency and the resulting coherent population trapping in the lower two states of the three-level $\Lambda$ system.

\par
Another pertinent point is the proper implementation of causality via the Kramers-Kronig relations \cite{sdgbook}, since all such NRI materials are essentially lossy. A great deal of research has gone into how to choose the proper signs of the real and imaginary parts of the electric and magnetic susceptibilities  and consequently the signs of both real and imaginary parts of the refractive index \cite{lakhtakia2004,stockman,kinsler2008}. Note that absorption plays a deciding role on the feasibility of the exotic effects like Pendry lensing and superresolution \cite{sdgepjap,subimal}. In our study, we pay due attention to the complications that may arise in molecular gaseous systems exhibiting gain.

Our work is motivated by the recent experiments \cite{zwierlein,ni2008,will2016} in producing cold polar molecules in the absolute ground state having permanent electric as well as magnetic dipole moment. In contrast to other proposed three-level schemes, our scheme differs in several aspects. First, our scheme does not involve any electronically excited state, and so our model has relatively stronger immunity to spontaneous decays, offering greater opportunity for coherent manipulation.  Second, because of microwave-dressing, in our model both electric and magnetic transitions are possible between the same two levels, unlike the $\Lambda$-type model \cite{oktel} where the magnetic transition occurs between the two lower levels and the electric one between the middle and the uppermost level. Third, our model appears to be realizable with currently available technology of cold polar molecules.   

\par
The remainder of the paper is organized in the following way. In section-\ref{sec:level1}, we discuss our model and obtain the expressions for  electric permittivity and magnetic permeability in terms of the  density matrix elements. In section-\ref{sec:level2}, we present our analytical and numerical results under certain parameters regime to obtain NIM in an ensemble of ground state polar molecules. Finally, we summarize the main findings of our study in conclusions.
    
\section{\label{sec:level1}THE MODEL}
We consider a three-level scheme of electronically ground-state polar molecules possessing both permanent electric and magnetic dipole moments, interacting with three laser fields. The configuration is depicted in  Fig.\ref{nalipotential} where the  states $\ket a $ and $\ket b$ have different parity while $\ket a $ and $\ket c $ have the same parity. These states are characterized by the vibrational quantum number $v$, the total angular momentum $J$ and its projection $M_J$ on space-fixed coordinate system.  By exploiting the permanent electric dipole moment, the state $\mid a \rangle$ can be coupled to another level $\ket b $ by a microwave laser L$_1$, resulting in two dressed states $\mid + \rangle$ and $\mid - \rangle$. Two infrared lasers L$_2$ and L$_3$ are applied to couple theses two dressed states to the level $\ket c $ forming a three-level $\Lambda$-type system as depicted in Fig.\ref{potential}(b).  Note that the laser coupling of the component $\mid b \rangle$ of the two dressed states to $\mid 
c \rangle$ occurs due to E1 transition  while the coupling of the component $\mid a \rangle$ to $\mid c \rangle$ is due to magnetic dipole transition (M1). 
Polar molecules, in particular ground-state diatomic polar molecules produced from cold atoms provide a new scope for the creation of a mixed-parity state \cite{zoe:prl:125:063401:2020}, opening a new perspective in creating metamaterial out of a polar molecular gas.

Let the electric field ${\mathbf E}_j$ of laser L$_j$ be expressed as ${\mathbf E}_j = E_{j0} \hat{\epsilon}_j \cos(\omega_j t)$ where $E_{j0}$ is the amplitude, $\hat{\epsilon}_j$ is the polarization and $\omega_j$ is the angular frequency of the $j$-th laser.  The Rabi coupling between  two molecular states (ro-vibrational states)  $\mid v, J, M_J \rangle $ and $ \mid v', J', M_J' \rangle $ is given by 
\bea 
\Omega_{j} =   -\frac{1}{\hbar} d E_{j0} \eta_{v'v} \langle J, M_J \mid \hat{d}\cdot \hat{\epsilon}_j \mid J' M_J' \rangle  
\eea
where $d$ is the magnitude of the dipole moment,  $\hat{d}$ is the unit vector along the dipole moment, $\eta_{v'v} = \int dr \psi_{v'}(r) \psi_{v}(r)$ is the Franck-Condon overlap integral between the two vibrational state $\psi_{v'}(r) = \langle r \mid v'\rangle$ and $\psi_{v}(r) = \langle r \mid v \rangle $ with $r$ being the interatomic separation.  The angular matrix element $ \langle J, M_J \mid \hat{d}\cdot \hat{\epsilon}_j \mid J' M_J' \rangle$ can be explicitly calculated using Wigner-Eckart theorem as discussed in the appendix-A. The resulting two dressed states  are given by
\begin{equation}
\ket {+}=\cos\theta \ket{a}+\sin\theta \ket{b}
\label{ds1}
\end{equation}
\begin{equation}
 \ket{-}=-\sin\theta \ket{a}+\cos\theta \ket{b}
 \label{ds2}
\end{equation}
where $\theta $ is the mixing angle defined by
 $\tan 2\theta = - \frac{\Omega_1}{\delta_1}$ with $\Omega_1$ being the Rabi coupling  and $\delta_1 = \omega_1 - \omega_{ba}$ the detuning of the laser frequency from the atomic frequency $\omega_{ba}$ for the transition $\ket b \leftrightarrow \ket a$. 
The two eigenvalues are $\hbar\tilde{\omega}_+=\frac{1}{2}\hbar(-\delta_1-\sqrt{\delta_1^2+\Omega_1^2})$ and $\hbar\tilde{\omega}_-=\frac{1}{2}\hbar(-\delta_1+\sqrt{\delta_1^2+\Omega_1^2})$. 
We restrict $\theta $ to lie in the range $0 < \theta < \pi/2$ implying $\delta_1 < 0$ and  $\tilde{\omega}_+<\tilde{\omega}_-$. So,  the two dressed states always  remain to be of mixed parity. For $\delta_1 = 0$, $\ket +$ and $\ket-$ reduce to a symmetric and anti-symmetric combinations, respectively, of the two bare states $\ket b$ and $\ket a$.

\par
Since magnetic dipole Rabi coupling is smaller than the electric one by two orders of magnitude, we consider only the electric dipole coupling in calculating the atomic density matrix. Nevertheless, the important point here is that we can generate magnetic response out of purely electrical couplings as originally suggested by Pendry \cite{pendry:1999}.  The Rabi couplings of the states $\ket +$ and $\ket -$ to $\ket c$ are $\Omega_
+ = \Omega_2~\sin \theta $ and $\Omega_- = \Omega_3~ \cos \theta$, respectively; where $\Omega_{2}\,(\Omega_3)$ is the Rabi coupling between the states $\ket b$ and $\ket c$ $(\ket a$ and $\ket c)$ due to the laser L$_2$ (L$_3$). Since we are considering all the three levels to lie in electronic ground state manifold,  the radiative damping or spontaneous decay rate $\gamma$ from the upper level $\ket{c}$  to the lower level $\ket b$ is quite small.   Since electric dipole coupling between $\ket c$ and $\ket a$ is forbidden by parity selection rule, there is no spontaneous emission from $\ket c$ to $\ket a$ due to electric dipole emission. Spontaneous emission linewidth is proportional to the square of the electric dipole matrix element. So,  the  linewidths for spontaneous emission from $\ket c$ to $\ket +$ and $\ket -$ are $\gamma_{+} = \gamma \sin^2 \theta $ and $\gamma_{-} = \gamma \cos^2 \theta $, respectively. In our model, we assume that the state $\ket-$ can decay to $\ket+ $ with small damping 
constant $\gamma_0$.   

\par
At probe frequency $\omega_2$, the electric dipole transition $(\ket b\leftrightarrow \ket c)$ and magnetic dipole transition $(\ket a\leftrightarrow \ket c)$ can occur simultaneously, producing electric  and magnetic polarization of the medium. The relative permittivity and  permeability are defined by  $  \epsilon_r=1+\chi_e$ and $
\mu_r=1+\chi_m$, respectively; where $\chi_e$ and $\chi_m$ are the electric and magnetic susceptibility, respectively.   The refractive index is $n = \pm \sqrt{\epsilon_r \mu_r}$. Here the negative sign (-) before $\sqrt{\epsilon_r \mu_r}$ is applicable only when  both $\epsilon_r$ and $\mu_r$ are real  and negative  or the phase velocity of the electromagnetic field  opposes the velocity of energy flow of the field \cite{kinsler2008,lakhtakia2007}. For a dissipative system, both $\epsilon_r$ and $\mu_r$ are complex. Let $\epsilon_r = \epsilon_r^{\prime} + i \epsilon_r^{\prime \prime}$ and $\mu_r = \mu_r^{\prime} + i \mu_r^{\prime \prime }$, where primed and double-primed quantities refer to the real and imaginary parts, respectively.  The real and imaginary parts of $\epsilon_r$ or $\mu_r$ are related to each other by causality conditions expressed by celebrated Kramers-Kronig relations.
According to the general criterion \cite{kinsler2008}, the refractive index $n$ becomes negative when the quantity $K_E = \epsilon_r^{\prime} |\mu_r| + \mu_r^{\prime} |\epsilon_r|$ is negative for passive systems.

 \par
 For a dilute medium  of molecular gases with number density $N$, the electric susceptibility  $\chi_e^{0}$ which describes the macroscopic dielectric property of the medium is related to the microscopic polarizability tensor $\alpha_e$ of a molecule   by $\chi_e^{0} = N \alpha_e/\epsilon_0$. The diluteness of a medium  may be characterized by $N \lambda^3  <\!<  1.0$, implying that the average number of atoms within a volume of cubic wavelength $ \lambda^3$ of the propagating wave is much less than unity. But for a dense gas of molecules or atoms, one has to take into account local-field effects by using the Lorenz-Lorenz relation between susceptibility or dielectric function and polarizability \cite{Jackson}, which in turn leads to the  Clausius-Mossoti equation. Early works  \cite{Bowden1993,GSA1996} have demonstrated nontrivial local field effects on spectral properties 
of dense media of two-level \cite{Bowden1993} and three-level atomic systems \cite{GSA1996}. In the context of creating negative refraction by laser manipulation of atomic gases, several authors \cite{oktel,zhang1,krowne3,budriga} have discussed the essential role of local field effects in terms of Lorenz-Lorenz formula \cite{Jackson} or Clausius-Mossoti equations. Orth {\it et al.} \cite{orth} have carried out a detailed study by numerically solving nonlinear density matrix equations to analyze local field effects on refractive index of a dense five-level atomic system.  In our approach we take into account the local field effects  in the spirit of Clausius-Mossoti equation.

\par
The Hamiltonian of our system is 
\bea 
\tilde{H} &=& \hbar \tilde{\omega}_+ \ket + \bra +  +  \hbar \tilde{\omega}_- \ket -\bra - 
+ \hbar \omega_c  \ket c \bra c  +\frac{\hbar}{2}\Omega_{+}e^{-i \omega_2 t} \ket{c} \bra{+} \\
&+& \frac{\hbar}{2}  \Omega_{-} e^{-i \omega_3 t} \ket{c}\bra{-} + {\rm H.c.}
\eea 
The density matrix equation is given by 
\bea 
 \frac{d \tilde{\rho}}{ d t} = - \frac{i}{\hbar} \left [\tilde{H}, \tilde{\rho} \right ] + {\mathcal L} \tilde{\rho}
\eea 
where ${\mathcal L}$ is a Liouvillian operator that describes dissipation of the system,
which can be recast into the set of coupled equations of motion for 
the density matrix elements. Explicitly, we have 
\begin{eqnarray}
\dot{\rho}_{--}&=&
-\frac{i}{2}\Omega_{-}[{\rho}_{c-}-{\rho}_{-c}]+\gamma_{-}{\rho}_{cc} - \gamma_0 {\rho}_{--}
\label{ro--}
\\
\dot{\rho}_{++}&=& -\frac{i}{2}\Omega_{+}[{\rho}_{c+}-{\rho}_{+c}]+\gamma_{+}{\rho}_{cc} + \gamma_0
{\rho}_{--} 
\label{ro++}
\\
\dot{\rho}_{cc}&=&\frac{i}{2}\Omega_{+}[{\rho}_{c+}-{\rho}_{+c}]+\frac{i}{2}\Omega_{-}[{\rho}_{c-}-{\rho}_{-c}]-\gamma{\rho}_{cc}\\
\dot{\rho}_{-+}&=& i[\delta_{+} - \delta_{-}]{\rho}_{-+}-\frac{i}{2}\Omega_{-}{\rho}_{c+}+\frac{i}{2}\Omega_{+}{\rho}_{-c} - \frac{\gamma_0}{2} {\rho}_{-+}\\
\dot{\rho}_{+-}&=& -i[\delta_{+} - \delta_{-}]{\rho}_{+-}+\frac{i}{2}\Omega_{-}{\rho}_{+c}-\frac{i}{2}\Omega_{+}{\rho}_{c-} - \frac{\gamma_0}{2} {\rho}_{+-}
\label{ro+-}
\\
\dot{\rho}_{+c}&=&-i\delta_{+} {\rho}_{+c}-\frac{i}{2}\Omega_{+}({\rho}_{cc}-{\rho}_{++})+\frac{i}{2}\Omega_{-}{\rho}_{+-}-\frac{\gamma}{2}{\rho}_{+c}
\label{ro+c}
\\
\dot{\rho}_{c+}&=& i\delta_{+} {\rho}_{c+}+\frac{i}{2}\Omega_{+}({\rho}_{cc}-{\rho}_{++})-\frac{i}{2}\Omega_{-}{\rho}_{-+}-\frac{\gamma }{2}{\rho}_{c+}\\
\dot{\rho}_{-c}&=&-i\delta_{-} {\rho}_{-c}+\frac{i}{2}\Omega_{+}{\rho}_{-+}-\frac{i}{2}\Omega_{-}({\rho}_{cc}-{\rho}_{--})-\frac{\gamma + \gamma_0}{2}{\rho}_{-c}
\label{ro-c}
\\
\dot{\rho}_{c-}&=& i \delta_{-} {\rho}_{c-}-\frac{i}{2}\Omega_{+}\hbar{\rho}_{+-}+\frac{i}{2}\Omega_{-}\hbar({\rho}_{cc}-{\rho}_{--})-\frac{\gamma + \gamma_0}{2}{\rho}_{c-}
\end{eqnarray}
where $\rho_{jj} = \tilde{\rho}_{jj}$ ($j=+,-,c$), $\rho_{+c} = e^{-i \omega_2 t} \tilde{\rho}_{+c}$, 
$\rho_{-c} = e^{-i \omega_3 t} \tilde{\rho}_{-c}$ and $\rho_{+-} = e^{-i (\omega_2 - \omega_3) t} \tilde{\rho}_{+-}$.
 Here, the two detuning parameters are defined as $\delta_+=\omega_2-(\omega_c-\tilde{\omega}_{+})$, $\delta_-=\omega_3-(\omega_c-\tilde{\omega}_{-})$.   Note that $\gamma = \gamma_+ + \gamma_{-}$ is the total decay rate of $\ket c$. 
The above density matrix equations are constrained by ${\rho}_{--}+{\rho}_{++}+{\rho}_{cc}=1$ and ${\rho}_{ij}={\rho}^*_{ji}$.
 Although we solve numerically the density matrix equation in steady state, for gaining insight we derive exact analytical expressions for the steady-state density elements in the appendix-B.

The transition dipole moment operator ${\mathbf d}$  acting between $\mid b \rangle$ and $\mid c\rangle$ is given by
${\mathbf d} =  d_{bc} \ket{b} \bra{c}  +  d_{cb} \ket{c} \bra{b} $. The average dipole moment of the molecule  is   $\langle {\mathbf d} \rangle =  {\rm Tr}[ {\mathbf d} \tilde{\rho} ] =  d_{bc} \rho_{cb} e^{-i \omega_2 t} +  d_{bc} \rho_{bc} e^{i \omega_2 t}$. Within the framework of linear response theory, we have $\langle {\mathbf d} \rangle = \frac{1}{2} \left [ \alpha_e E_{20} e^{-i \omega_2 t} + \alpha_e^* E_{20}^{*} e^{i \omega_2 t} \right ]$, where $E_{20}$ is electric field amplitude of the probe laser $L_2$. Thus  we have
\begin{eqnarray}
\alpha_e=\frac{2d_{bc}\rho_{cb}}{E_{20}}.
\end{eqnarray}
 In the absence of any local field effect, the polarization of the medium is  ${\mathbf P}^{(0)} = N \langle  {\mathbf d} \rangle = \frac{\epsilon_0}{2} \left [ \chi_e^{0} E_{20} e^{-i\omega_2 t}
+   \chi_e^{0*} E_{20}^{*} e^{i \omega_2 t} \right ]$, implying  $\chi_e^0 = N \alpha_e/\epsilon_0$.  
By making use of Clausius-Mossoti relation, the electric susceptibility  $\chi_e$ for a dense medium is given by
\bea
\chi_e = \chi_e^0 \left [ 1 - \frac{\chi_e^0}{3 } \right ]^{-1}
\eea


To evaluate $\mu_r$, we need to calculate the magnetic dipole moment matrix element $\mu_{ca}$ between $\mid c \rangle$ and $\mid a \rangle$.  
 Following the Ref.\cite{oktel}, within linear response theory, this can be accomplished by expanding the matrix element $\rho_{ca}$ as a function of field amplitude $E_{20}$ and retaining only the first order term. This procedure leads to an expression for the magnetization ${\mathbf M}$  \cite{oktel}
\bea 
{\mathbf M} = -i  N \beta(\omega_2) c \left (1+ \frac{\chi_e}{3} \right ) {\mathbf B}
\eea 
where 
\bea 
\beta(\omega_2) = 2\mu_{ac} \frac{\partial \rho_{ca}}{\partial E_{20}}
\eea 
is a function of the probe frequency $\omega_2$. Then from the expressions ${\mathbf B} = \mu_0 \left ({\mathbf H} + {\mathbf M} \right)$ 
and ${\mathbf B} = \mu {\mathbf H}$, one obtains 

\bea 
\mu_r = \frac{1} { 1 + iN \mu_0 \beta c [1 + \chi_e/3]} \label{mu}
\eea 

Here $\epsilon_r$ and $\mu_r$ depends on the bare-state density matrix elements which are related to the dressed-state density elements $\rho_{+c}$ and $\rho_{-c}$ by
\begin{eqnarray}
 \rho_{bc}=\sin\theta\rho_{+c}+\cos\theta\rho_{-c}\label{rho_bc}\\
  \rho_{ac}=\cos\theta\rho_{+c}-\sin\theta\rho_{-c}\label{rho_ac}
\end{eqnarray}
which can be readily obtained by using Eqs. \ref{ds1} and \ref{ds2}.

\section{\label{sec:level2}Results and discussions}
\begin{figure}
\center
\includegraphics[width=0.8\linewidth]{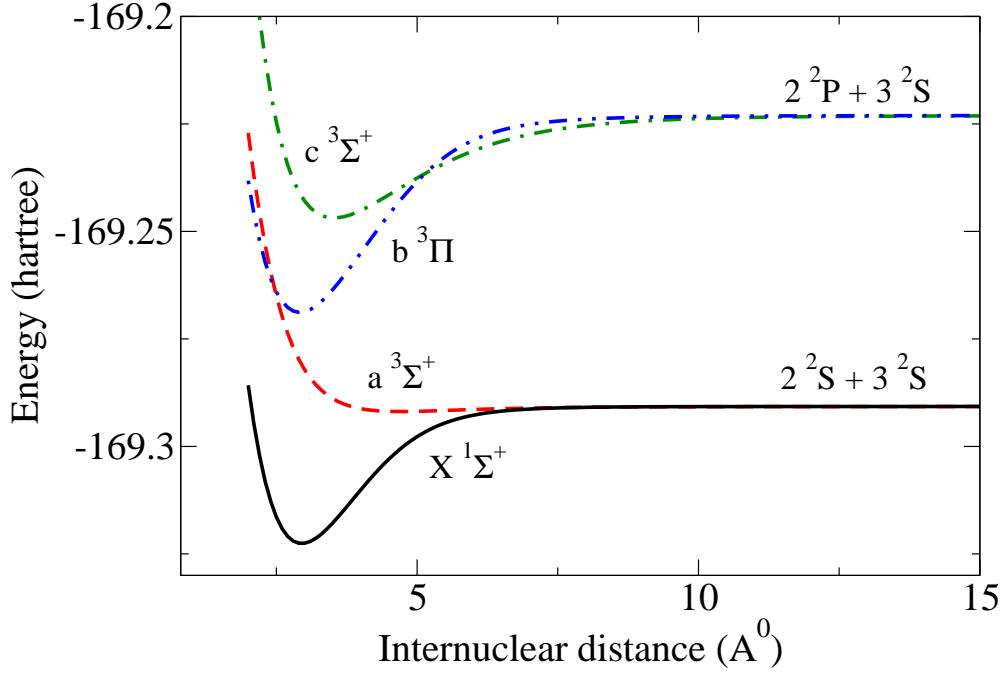}
\caption{The adiabatic potential energy curves of the four lowest electronic molecular states of NaLi system. In the large separation limit, the ground-state potentials of $X ^1\Sigma^+$ and $a ^3\Sigma^+$ electronic states go as $2\,\,^2$S$\,+\,3\,\,^2$S while the excited-state potentials go as  $2\,\,^2$P$\,+\,3\,\,^2$S.}
\label{nalipotential}
\end{figure}

For numerical illustration, we consider the long-lived diatomic polar molecule  $^{23}$Na$^{6}$Li molecule in spin-triplet electronic ground state a$^3\Sigma^{+}$ state which has been experimentally produced three years back by Rvachov {\it et al.} \cite{zwierlein}. The ground singlet, triplet and two excited potentials are shown in Fig.\ref{nalipotential}.  This molecule has a magnetic dipole moment of 2$\mu_B$ ($\mu_B$ is Bohr magneton) and a permanent electric dipole moment of 0.2D (D stands for Debye). In the electronic ground state, the total angular momentum is $\mathbf{J} =  {\mathbf N} + 
{\mathbf S}$ where  $S$  is the   spin angular momentum  and $N$  the rotational angular momentum of the internuclear motion. For simplicity, we here ignore molecular hyperfine interaction.  The parity of this spin-triplet  molecular state is given by $(-1)^N$ \cite{Herzberg:book}. Each $N$ level are splits into the corresponding $J$ levels due to spin-axis interaction  \cite{Herzberg:book}.  
In the experiment of Ref. \cite{zwierlein}, the molecule is  prepared  in the absolute ground state, meaning all the quantum numbers of the state are minimum, that is, $N=0$, $J=1$ and the vibrational quantum number  $v=0$.  Thus we can justifiably assume  that our model molecular system is initially prepared in the  state $\ket{a} \equiv  \ket{v=0,N=1, J=1}$.   Let us suppose that the three magnetic sub-levels of $J=1$ of $\mid a \rangle$ are split by a biased magnetic field. Let the microwave laser L$_1$ be linearly polarized along the magnetic field direction and tuned between the transition $\mid a (v=0, N=1, J=1, M_J=1)\rangle \rightarrow \mid b(v=0, N=2, J=1, M_J=1)\rangle$. Let the two dressed states be  coupled to the level $\ket c$ having $v=2$, $N=1$, $J=2$ and $M_J = 0$ with two circularly polarized infrared lasers L$_2$ and L$_3$. The fact  that the electric dipole moment matrix element for transitions between the states $\mid a \rangle$ and $\mid b \rangle$ and that between $\mid b \rangle$ and $\mid c \rangle$ are nonzero, while there exists nonzero  magnetic dipole moment matrix element between $\mid a \rangle$ and $\mid c \rangle$  is illustrated with explicit calculations in the appendix-A. 

It is well known that, for $\gamma_0 = 0$ and  $\Delta =\delta_{+}-\delta_{-}= 0$ (two-photon resonance), there is dark-state resonance at which ${\rho}_{cc} = 0$ or equivalently ${\rm Im}[{\rho}_{+c}] = 0$ and ${\rm Im}[{\rho}_{-c}] = 0$ meaning that all the absorption absolutely vanishes. All the population is then coherently shared  by the two lower states $\ket +$ and $\ket -$ with population ratio given by ${\rho}_{--}/{\rho}_{++} = w_-/w_+ = r = \Omega_{+}^2/\Omega_{-}^2$. This happens due to the quantum interference between optical transition pathways connecting the uppermost state $\ket {c}$ to the lower states $\ket {+}$ and $\ket {-}$. It is clear from the expression (\ref{eqnpcstnew}) that at such dark-state resonance conditions, ${\rho}_{+c} = 0$ implying that both the real and imaginary parts vanish. The question here is whether there exists a dispersive regime close to the dark-state resonance or quasi-dark state resonance ($\gamma_0 <\!< \gamma$, but $\gamma_0 \ne 0$) where the desirable negative index of refraction is possible.

\par
We have derived exact analytical expressions for the density matrix element ${\rho}_{+c}$ given by the equation (\ref{eqnpcstnew}) in the appendix-B. Setting $\gamma_0 =  0$,   ${\rm Re}[{\rho_{+c}}]$ can be expressed in a compact form 
\begin{eqnarray}
 {\rm Re}[{\rho_{+c}}] = F_{+}\left [ A + B (r - r_d) \right ]
 \label{repccompact}
\end{eqnarray}
where
\begin{eqnarray}
 F_{+} = \frac{w_{+}  \Omega_{+}} {8  |\kappa_{+}|^2 \left [ (\Delta - \Delta_{{\rm shift}})^2  +  (\Gamma/2)^2 \right ]} 
\end{eqnarray}
\begin{eqnarray}
 A =  \left [ 4 \delta_{+} \Delta     \left (\Delta -  \Delta_{{\rm shift}} \right ) + \Delta \gamma_{+} \Gamma  \right ]
  \label{eqna}
\end{eqnarray}
\begin{eqnarray}
 B =  \frac{\Omega_{-}^2}{|\kappa_{-}|^2} \left [  \left (\Delta -  \Delta_{{\rm shift}} \right ) \left ( \delta_{+} \delta_{-} +  \gamma_{-} \gamma_{+}/4 \right ) + \Gamma/4 \left ( \delta_{-}\gamma_{+}   -  \delta_{+}  \gamma_{-} \right) \right ]
\end{eqnarray}
Here $r_d = \Omega_{+}^2/\Omega_{-}^2$ and 
\begin{equation}
 \Delta_{{\rm shift}} = \frac{1}{4} \left [ \frac{\Omega_{-}^2 \delta_{+}}{|\kappa_{+}|^2} - \frac{\Omega_{+}^2 \delta_{-}}{|\kappa_{-}|^2} \right ]
\end{equation} 
\begin{equation}
 \Gamma = \gamma_0 + \frac{1}{2} \left [ \frac{\Omega_{-}^2 \gamma_{+}}{|\kappa_{+}|^2} + \frac{\Omega_{+}^2 \gamma_{-}}{|\kappa_{-}|^2} \right ]
\end{equation}
The structure of equations \ref{eqnpcstnew1} and \ref{eqnmcstnew1} suggest that ${\rho}_{+c}$ can be obtained from the expression of ${\rho}_{-c}$ by changing all subscripts \textquoteleft+\textquoteright$ \rightarrow$ \textquoteleft-\textquoteright and vice versa. So, substituting these expressions in equations \ref{rho_bc} and \ref{rho_ac}, we obtain exact analytical expressions for $\rho_{bc}$ and $\rho_{ac}$. 
 
 \begin{figure}
 	\center
 	\includegraphics[width=0.9\linewidth]{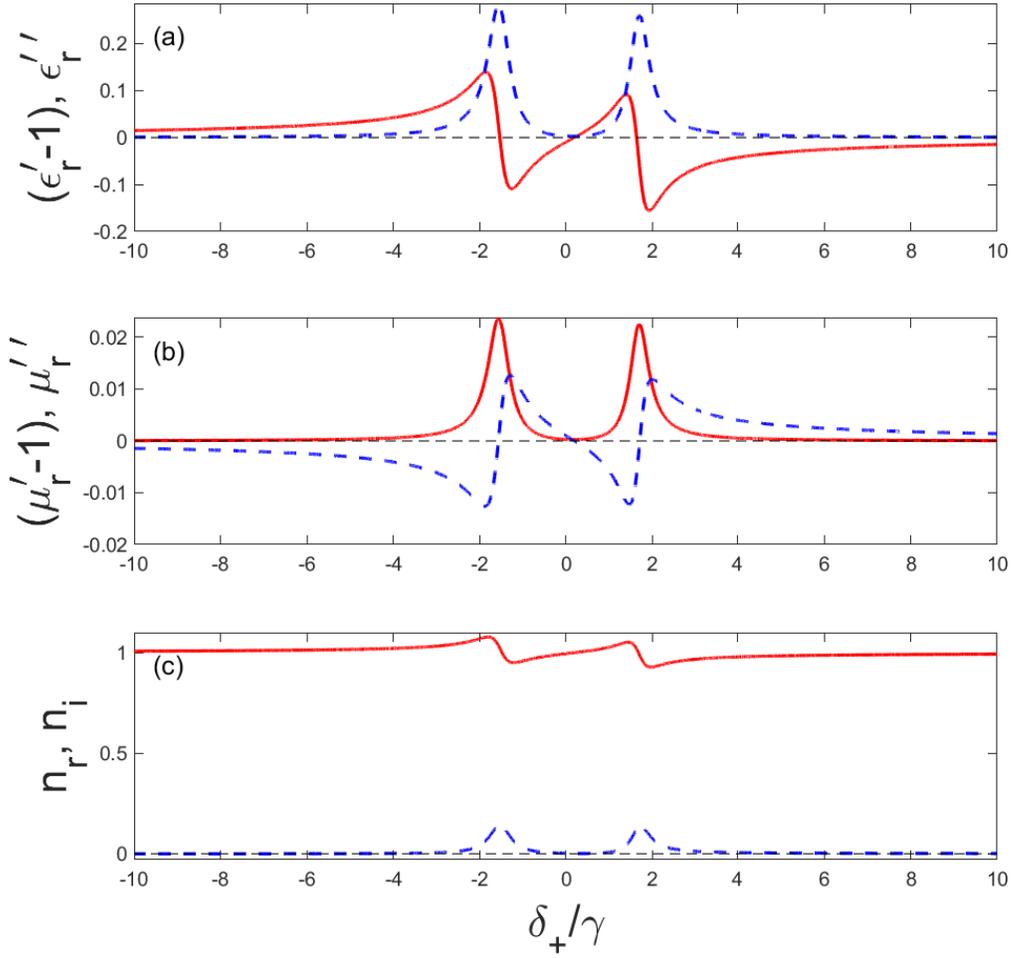}
 	\caption{(a) and (b):  The real parts $(\epsilon_r'-1)$ and $(\mu_r'-1)$ (red, solid) and the imaginary parts $\epsilon_r{''}$ and $\mu_r^{''}$
 		(blue, dotted) of the electric and the magnetic susceptibility for a low density with $N=4\times10^{14}$ cm$^{-3}$ as a function of dimensionless detuning $\delta_+/\gamma$. (c): The real part $n_r$ and the imaginary part $n_i$ of the refractive index $n$ as a function $\delta_+/\gamma$. ; The other parameters are $\gamma_0=0.1\gamma$, $|\Omega_2|=0.2\gamma$, $|\Omega_{3}|=3.6\gamma$, $\delta_{-}=0.2\gamma, \theta=\pi/7$   .}
 	\label{figure:LD}
 \end{figure}

\begin{figure}
\center
\includegraphics[width=0.9\linewidth]{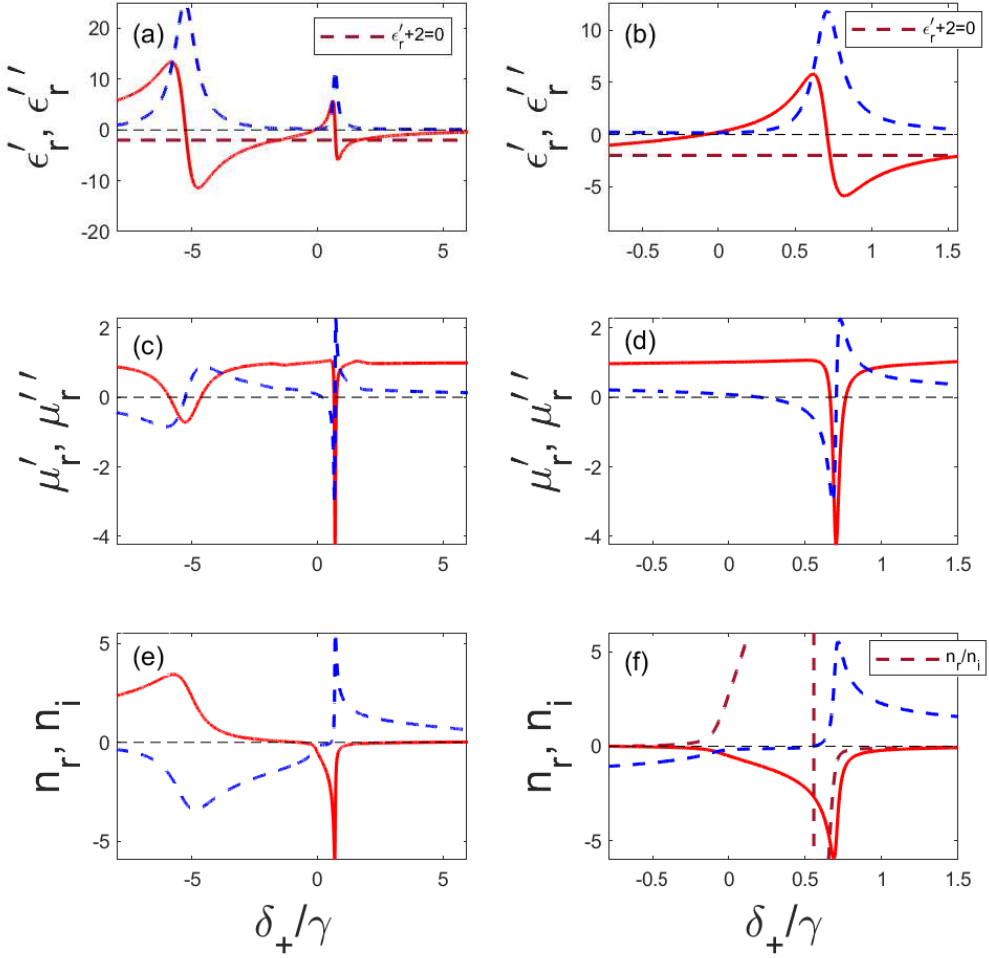}
\caption{(a) and (c):  The real parts $\epsilon_r'$ and $\mu_r'$ (red, solid) and the imaginary parts $\epsilon_r{''}$ and $\mu_r^{''}$
(blue, dotted) of $\epsilon_r$ and $\mu_r$  as a function of dimensionless detuning $\delta_+/\gamma$. (e): The real part $n_r$ and the imaginary part $n_i$ of the refractive index $n$ as a function $\delta_+/\gamma$. (b) and (d): $\epsilon_r$ and $\mu_r$ as a function of $\delta_{+}$ zoomed around the two-photon resonance point. (f) Zoomed version of the subplot (e) along with the ratio $n_r/n_i$; The other parameters are $\gamma_0=0.1\gamma$, $|\Omega_2|=0.2\gamma$, $|\Omega_{3}|=3.6\gamma$, $\delta_{-}=0.2\gamma, \theta=\pi/7$ and the number density, $N=4\times10^{16}$ cm$^{-3}$.}
\label{figure1}
\end{figure}

\begin{figure}
	\center
	\includegraphics[width=0.9\linewidth]{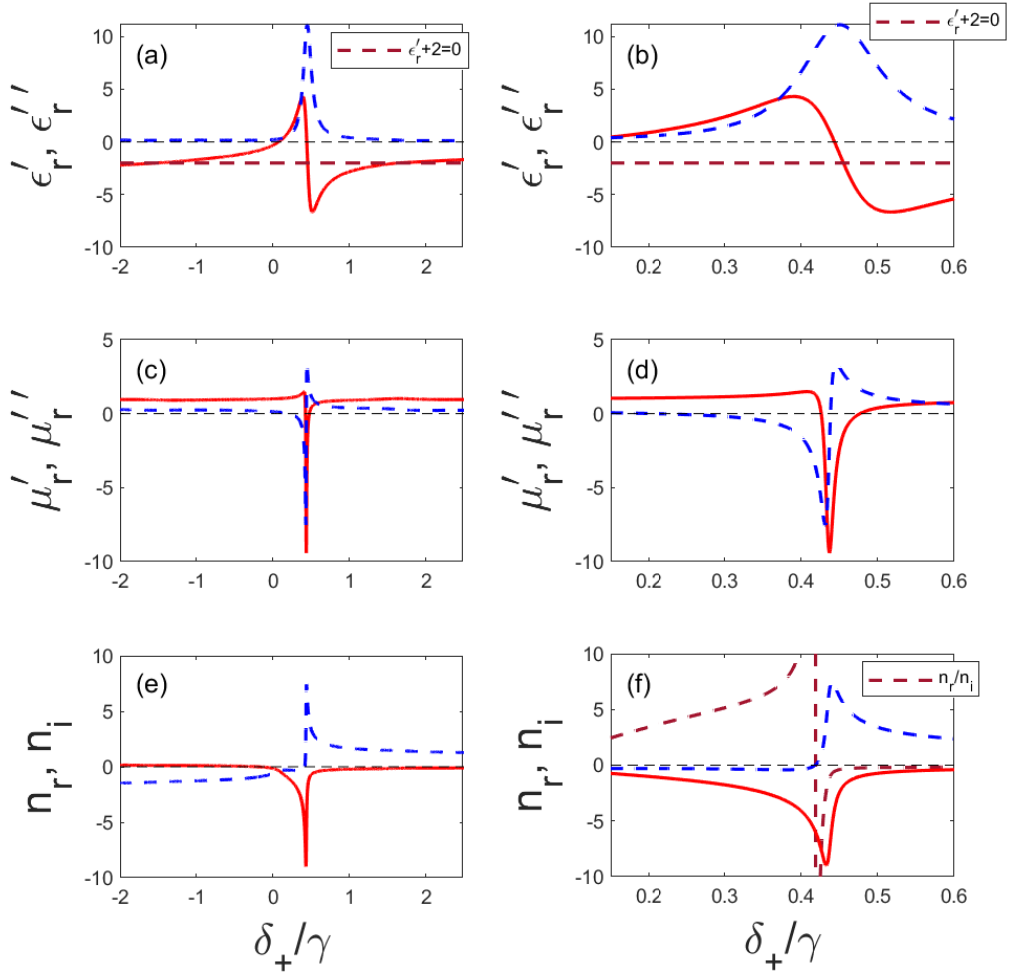}
	\caption{Same as in Fig.\ref{figure1} but for a higher number density, $N=9\times10^{16}$ cm$^{-3}$.}
	\label{figure2}
\end{figure}

\begin{figure}
	\center
	\includegraphics[width=0.9\linewidth]{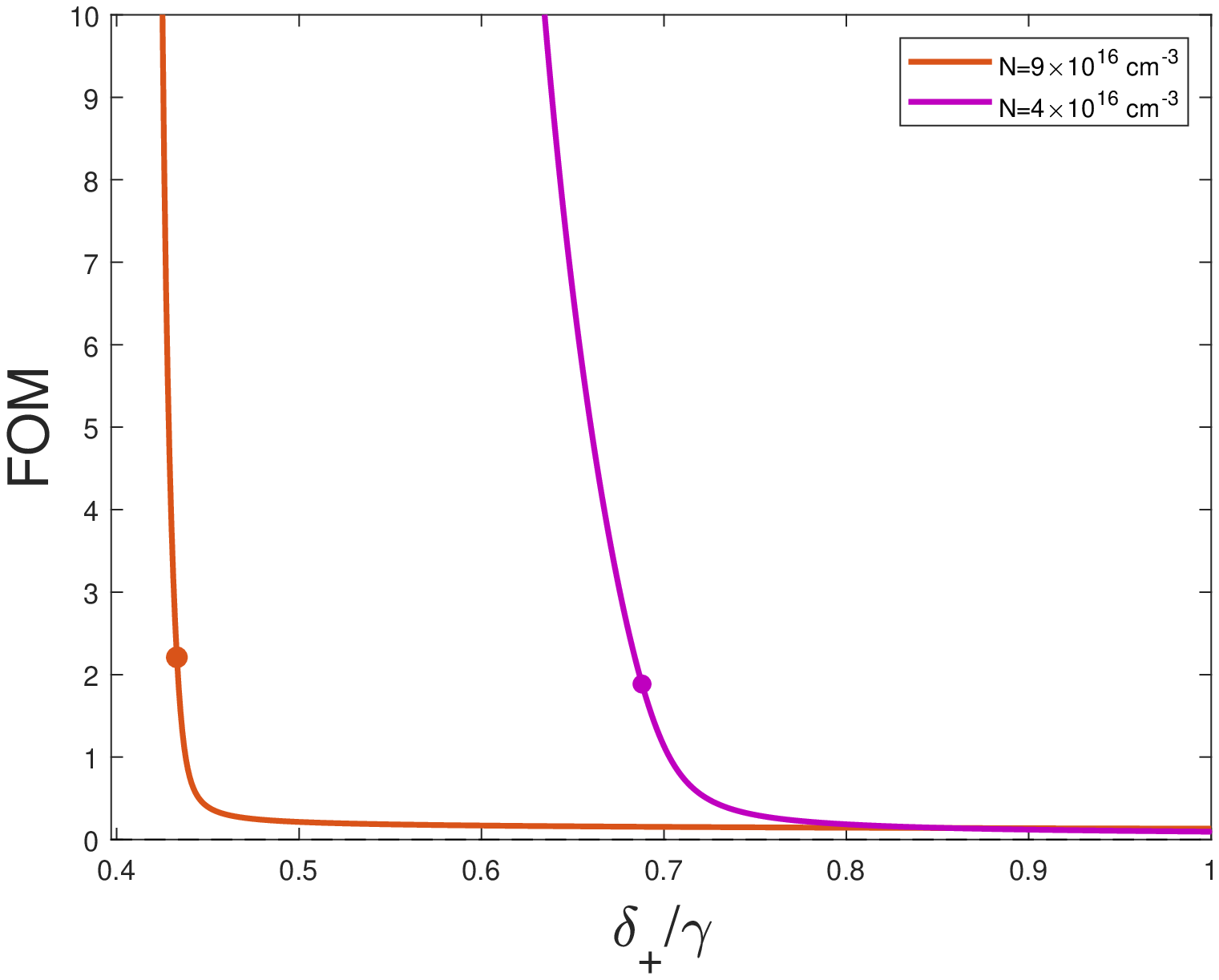}
	\caption{The figure of merit (FOM) for the results shown in  Fig.\ref{figure1} and \ref{figure2}. The values of FOM at the minimum  of the 
	negative refractive indexes (the minimum values of $n_r$ are -5.93 and - 8.9  for $N = 4 \times 10^{16}$ and $N = 9 \times 10^{16}$, respectively) are indicated by dots.}
	\label{FOM}
\end{figure}

\par
Having discussed the analytical results, we next turn our attention to numerical results which do corroborate our analytical findings.  
Making use of the numerical solutions of density matrix equations in steady state, we calculate the complex relative permittivity $\epsilon_r$, the magnetic permeability $\mu_r$ and the index of refraction $n$. 
We use  the magnitude of dipole moment $|d| = d_{cb}=0.4$ Debye, magnetic moment $\mu_{ca}=2 \mu_B$, mixing angle $\theta=\pi/7$. The  values of permanent dipole and magnetic moments correspond to polar molecule  NaLi in triplet ground state.
In our calculations, we scale all the frequency parameters by spontaneous linewidth $\gamma$ for transition $\ket c\rightarrow \ket a $. The complex refractive index is $n = n_r + i n_i$, where $n_r$ and $n_i$ are the real and imaginary parts of $n$. The figure of merit (FOM) of a negative index material can be defined as FOM$=|{{\rm Re}}[n]/{\rm Im}[n]|$ that  characterizes the performance of a negative index material.

{To begin with we start with a dilute gas ensemble, where local field effects are not significant (see  Fig.\ref{figure:LD}). As can be seen from the figure the relative permittivity and permeability do not change appreciably from unity, and  over the frequeny regime studied there is no negative refraction. A simple taylor series expansion of Eqn.(\ref{mu}) shows that $Im(\mu_r)$ follows $Re(\rho_{ca})$, resulting in its dispersive nature. This interchange of roles of the real and imaginary parts of $\mu_r$ can be traced to the circular polarization of light, which introduces a $\pi/2$ phase difference between the electric and magnetic responses. This causes the imaginary part of $\mu$ to be negative signifying gain in the magnetic transition. Note, however, that overall the  system remains passive ($n_i> 0$) over the spectral domain.}

	\par
	 In contrast, for higher densities and hence unavoidable local field effects, the situation  changes drastically  leading to qualitative and quantitative differences. In order to show this we present results for number density $N= 4 \times 10^{16}$ cm$^{-3}$ in Fig.\ref{figure1}. 
In the subplots (a) and (c) of this figure, we show the variation of the real parts $\epsilon_r^{\prime}$ and $\mu_r^{\prime}$ (red, solid) and imaginary parts  $\epsilon_r^{\prime \prime} $ and $\mu_r^{\prime \prime }$ (blue, dotted) as a function of $\delta_{+}$ while the subplot (e)  of the same figure shows the variation of $n_r$ and $n_i$. In the right panels, we have plotted the same as in the left panel, but the negative refraction domain is zoomed for better clarity in viewing. These plots clearly demonstrate the  frequency regime of NRI. 

In Fig.\ref{figure2}, we have plotted the same as in Fig.\ref{figure1} but for a higher value of number density $N=9\times10^{16}$ cm$^{-3}$. Compared to Fig.\ref{figure1}, we find that an increased density leads to a smaller width of negative refraction regime. The maximum of absorption (${\rm Im}[\epsilon_r]$) shows red-shift with the increase of number density in conformity with earlier studies. Also, we find that the minimum and maximum of both the dispersion curves (${\rm Re}[\epsilon_r]$ and  ${\rm Re}[\mu_r]$) exhibit red-shifts due to increase in the number density. However, the FOM at the minimum of the negative index ($n_r$)  as shown in Fig. \ref{FOM} increases with the number density.

\par
It may be noted that the resonances in Fig.\ref{figure1}(a) occur near Re$[\epsilon_r]+2\sim0$, which is reminiscent of the localized plasmon resonances of small metal particles \cite{sdgbook,Bohren:Book,Mayer2011}. As in the plasmonic equivalent, here also one has large local field enhancements, which play an important role for left handed response.
\par
{It is pertinent to comment on the origin of gain in the overall response of the medium heralded by the negative imaginary part of the refractive index. We have verified that the gain in the medium is not a consequence of population inversion neither in the bare nor in the dressed systems. In fact, for the said system parameters there is no inversion in bare or dressed basis. Partly the gain can be traced to the interchange of the dispersive and absorptive behavior like in the case of low densities. However, for larger densities, the situation is less amenable to analysis since the Taylor series expansion is no longer valid and the $Im(\mu_r)$ can no longer be linked directly to $Re(\rho_{ca})$. In passing, it is worthwhile to remark that our three-level Lambda system consisting of $\ket{+},\,\ket{-}$, and  $\ket{c}$ is fundamentally different from a standard Lambda system in that the lower two states $\ket{+}$ and $\ket{-}$ are of mixed parity states. Here microwave dressing creates additional atomic coherence as can be evidenced from {Eqs. \ref{rho_bc} and \ref{rho_ac}} which show that the atomic coherences $\rho_{bc}$ and $\rho_{ac}$ are superpositions of $\rho_{+c}$ and  $\rho_{-c}$, and the superposition is controlled by the microwave mixing angle $\theta$. So, considering that the frequency dependence of $\rho_{+c}$ and $\rho_{-c}$ will follow those of a standard lambda system, $\rho_{bc}$ and $\rho_{ac}$ will deviate from those of standard one because of this coherent superposition. We have found an optimum value of $\theta=\pi/7$ for obtaining negative index.}

\par
For a passive medium with positive imaginary parts of both $\epsilon(\omega)$ and $\mu(\omega)$ the criterion for negative refractive index and left handed behavior is now well understood \cite{lakhtakia2004,stockman,kinsler2008}. However, for a gain medium the situation can be complex. In fact, there is a lot of debate about proper implementation of the Kramers-Kronig relation for a gain medium \cite{Skaar2008,Baranov2015}. As shown by Skaar  \cite{Skaar2008}, the situation gets worse when analytic properties of $\epsilon(\omega)\mu(\omega)$ are lost in the upper half complex plane as in the case of an inverted Lorentzian response \cite{Skaar2006a,Skaar2006b}. In case of a molecular gas (as in our case) the product $\epsilon\mu$ has a branch point in the upper half complex plane leading to the loss of analyticity. Hence, none of the accepted criteria for negative refraction in passive media are applicable here. We extract $n$ as a function of detuning by demanding continuity of the response and the fact that 
$n(\delta_+)\rightarrow+1$ as $\delta_+\rightarrow\infty$.
The same results can be obtained using the explicit expressions for the real and imaginary parts of the refractive index $n$ \cite{lakhtakia2004}, and have been widely used in literature \cite{zubairy}.
 \begin{eqnarray}
   n_r &=& \pm \frac{1}{\sqrt 2} (|\epsilon_r||\mu_r|+\mu_r'\epsilon_r'-\mu_r''\epsilon_r'')^{1/2}\\
   n_i &=& \pm\frac{1}{\sqrt 2} \frac{\mu_r'' \epsilon_r'+\mu_r'\epsilon_r''}{(|\epsilon_r||\mu_r|+\mu_r'\epsilon_r'-\mu_r''\epsilon_r'')^{1/2}}
 \end{eqnarray} 
 {Again the sign needs to be picked demanding the continuity of both the real and imaginary parts of the response and the condition at $\delta_+\rightarrow\infty$.}

Before ending this section, it may be worthwhile to comment about the input parameters chosen for our numerical calculations. All frequency parameters are scaled by $\gamma$. For NaLi molecule, the lifetime of its absolute ground-state ($v=0$) is reported to be 4.6 s. The lifetime of the vibrational state $v=2$ is unknown. To be realistic, if we assume it to be of the order of $0.1$ or $0.01$ second, implying $\gamma\sim10$ Hz or $\gamma\sim100$ Hz, then from the value of $|\Omega_{2}|$ used for the plots, we estimate the intensity of the probe laser will be of the order of $1$ mWcm$^{-2}$ while that of the control laser will be the order of $100$ mWcm$^{-2}$. As these parameters are quite realistic, we hope our model will be realizable with currently available laser manipulation techniques of cold molecular gases.

\section{\label{sec:level3}Conclusions}
We have proposed a scheme to realize negative refraction in an ensemble of gaseous polar molecules having both levels of ground vibrational states. Along with two more IR lasers coupling the dressed states to the excited state enables one to extract both electric and magnetic response from the system. The underlying EIT level scheme ($\Lambda$ system) and the associated low losses and flexibility for dispersion control are shown to lead to  NRI behavior.  We report the possibility of having a FOM  as high as 10. Our calculations are based on a density matrix approach with numerical and exact analytical results for the steady states. The permittivity and permeability calculations are carried out keeping in view the complexities that might arise in gain media.

Of late, cold polar molecules emerge as a novel system for applications in coherent optics. In our model, micro-wave-dressed states of the two rotational levels play an essential role in generating both electric and magnetic response at the same frequency. Field-dressing of atomic levels have been used over the years for coherent manipulation of the optical properties of the atoms. In a recent work by Rempe's group \cite{boas2020}, atomic dressed states are employed to build a multi-level atomic scheme to control nonlinear response of the system at a quantum level. Since atoms do not have permanent electric dipole moment, fitting  our model into an atomic system seems to be unlikely. However, our model may be extended to include electronically excited manifold of cold polar molecules to obtain NRI at higher frequencies beyond infrared domain.  
\vspace{0.5cm} 

\noindent
{\bf Acknowledgment} \\
Dibyendu Sardar is thankful to Council of Scientific and Industrial Research  (CSIR), Government of India,  for a  support. Sauvik Roy is thankful to the Department of Science and Technology (DST), Government of India for INSPIRE fellowship.

\appendix 
\section{Calculation of molecular dipole matrix elements}

Here we calculate the tensor operator matrix elements associated with the electric and magnetic dipole matrix elements between  the angular momentum states of NaLi molecule in its 
spin-triplet electronic ground state a$^{3}\Sigma^+$ as discussed in section \ref{sec:level2}. Since a$^{3}\Sigma^+$ belongs to Hund's case (b), an appropriate angular momentum state for this case is $\mid  \Lambda; N, S, J, M_J \rangle $ where $\Lambda$ is the projection of electronic orbital angular momentum onto the internuclear axis of the molecule. Let us first consider the product of the operators $\hat{d}\cdot \hat{\epsilon} = \sum_{q=0,\pm 1} T^{(1)}_q(\hat{d}) T^{(1)*}_q(\hat{\epsilon})$ defined in the space-fixed coordinate system, where $T^{(1)}(\hat{d})$ and $T^{(1)}(\hat{\epsilon})$ are the spherical tensor operators of rank 1. The operator  $T^{(1)}(\hat{d})$ acts on the molecular states which have well-defined angular momentum with respect to a molecular body-fixed symmetry axis which, in the present context, is the internuclear axis. We therefore need to transform  $T^{(1)}(\hat{d})$  into the body-fixed coordinate frame. This is done using Wigner $D$-matrices. So, we have 
\bea 
 T^{(1)}_q(\hat{d}) = \sum_{q'}  T^{(1)}_{q'} (\hat{d}) D_{q' q}(\Theta, \Phi, \Phi')
 \eea 
where $\Theta, \Phi, \Phi'$ are the Euler angles for the transformation between space and body-fixed coordinate systems. For polarized light, only one $q$ value will contribute to the matrix element. For linearly polarized light, $q=0$ while for circularly polarized light $q= +1$ or $q=-1$. Since  the molecule has a definite $\Lambda$ value (in our case, $\Lambda = 0$), only one $q'$ value ($q'=0$) will contribute to the matrix element. Let us consider the matrix element between the two states $\mid  \Lambda; N, S, J, M_J \rangle $ and   $\mid  \Lambda; N', S, J', M_J' \rangle $. Following the standard angular momentum algebra applicable to the diatomic molecules \cite{Brown:Book}, we find 
\bea
\langle \Lambda; N, S, J, M_J \mid  T^{(1)}_{q'} (\hat{d}) D_{q' q}(\Theta, \Phi, \Phi') \mid  \Lambda; N', S, J', M_J' \rangle  \nonumber \\
= (-1)^{J + J' + N + S - M_J } 
\sqrt{(2 J'+1)(2J+1)} \left (
 {\begin{array}{ccc}
 J & 1  & J' \\
 -M_J & q & M_J' \\
\end{array} }
\right ) \nonumber \\
\times  \left \{ 
{\begin{array}{ccc}
  J & N & S \\
  N' & J' & 1 \\ 
 \end{array} }
 \right \}
 (1)^{N-\Lambda} \sqrt{(2N+1)(2N'+1)} \left (
 {\begin{array}{ccc}
  N & 1 & N' \\
  -\Lambda & q' & \Lambda \\
 \end{array}}
 \right )
\eea
where $()$ and $\{\}$ are the Wigner 3j and 6j symbols, respectively. 
Since, in the present context, $\Lambda = \Lambda' = 0$ and $q'=0$, the second 3j symbol involving $N$ and $N'$ will be nonzero if $\Delta N = N-N' = \pm 1 $. For $q=0$ and $J' = J$, the first 3j symbol appearing in the above expression will be nonzero if $M_J = M_J' \ne 0$. For $q=\pm 1$, the 3j symbol will be nonzero for $\Delta J  = (J - J') = \pm 1$. So, both the matrix elements of electric dipole moment between $\ket{b} \equiv \ket{\Lambda = 0; N=2,  S=1, J=1, M_J=1} $ and $\ket{c}  \equiv  \ket{\Lambda=0; N'=1, S=1,  J'=2, M_J'=0}  $ for circularly polarized lasers and between $\ket{a} \equiv \ket{\Lambda = 0; N=1,  S=1, J=1, M_J=1} $ and $\mid b \rangle $  for linearly polarized microwave laser L$_1$ will be nonzero. 

 Let us next examine the conditions for which the magnetic dipole matrix element between the states $\mid  \Lambda; N, S, J, M_J \rangle $ and $\mid  \Lambda; N', S, J', M_J' \rangle $ is nonzero.  The  magnetic dipole interaction of a molecule with the magnetic field ${\mathbf B}$ of a laser  is $H_B = - {\vec{\mu}} \cdot {\mathbf B} = - {\mu_B} \left [ g_s {\mathbf S} - {\mathbf L} \right] \cdot {\mathbf B}$, where ${\vec{ \mu}}$ is the magnetic dipole moment, $\mu_B$ is the Bohr magneton,  $g_s$ is the spin gyromagnetic ratio and ${\mathbf L}$ is the  electronic orbital angular momentum of the molecule.  For a molecule in $\Sigma$ state, the matrix element of ${\mathbf L} \cdot {\mathbf B}$ vanishes. So, we need to calculate the matrix element of $-\mu_B g_s {\mathbf S} \cdot {\mathbf B}$ only. One can express this in the form $H_B = -\mu_B g_s B \sum_{q} T^{(1)}_{q}(\hat{\mu}) T^{(1)*}_{q}(\hat{B})$, where $\hat{\mu}$ and $\hat{B}$ are the unit vectors pointing towards ${\vec{\mu}}$ and ${\mathbf B}$. Explicitly we have  
\bea
\langle \Lambda; N, S, J, M_J \mid  T^{(1)}_{q} (\hat{\mu}) \mid  \Lambda; N', S, J', M_J' \rangle  \nonumber \\
= \delta_{N N'} (-1)^{J + J' + N + S - M_J } 
\sqrt{(2 J'+1)(2J+1)} \left (
 {\begin{array}{ccc}
 J & 1  & J' \\
 -M_J & q & M_J' \\
\end{array} }
\right ) \nonumber \\
\times  \left \{ 
{\begin{array}{ccc}
  J & S & N \\
  S & J' & 1 \\ 
 \end{array} }
 \right \}
  \sqrt{S(S+1)(2S+1)} 
\eea 
So, the above matrix element will be nonzero if $\Delta N = 0$.  
 So, for our chosen angular momentum states of $\mid a \rangle \equiv \mid  \Lambda=0; N=1, S=1, J=1, M_J =1 \rangle $ and $\mid c \rangle \equiv \mid  \Lambda=0; N=1, S=1, J=2, M_J = 0 \rangle $ the magnetic dipole moment matrix element will be nonzero for circularly polarized lasers.  

\section{\label{sec:level4}Steady-state solution of the density matrix} 

In the steady-state, the Eqs. \ref{ro--} and \ref{ro++} result in 
\begin{eqnarray}
{\rho}_{cc} &=& \frac{1}{\gamma_{-}} \left \{
\Omega_{-} {\rm Im}[{\rho}_{-c}] + \gamma_0 {\rho}_{--} \right \}  = \frac{1}{\gamma_{+}} \left \{ \Omega_{+}{\rm Im}[{\rho}_{+c}] - \gamma_0 {\rho}_{--} \right \}
\label{eqncc} 
\end{eqnarray}
 The Eqs. \ref{ro+c} and \ref{ro-c} in the steady-state yield 
\begin{eqnarray}
 {\rho}_{+c} = \frac{1}{2 (\delta_{+} - i \frac{\gamma }{2})} \left [  - \Omega_{+}({\rho}_{cc}-{\rho}_{++})+ \Omega_{-}{\rho}_{+-} \right ]
 \label{eqnpcst}
\end{eqnarray}
and 
\begin{eqnarray}
 {\rho}_{-c} = \frac{1}{2 (\delta_{-} - i \frac{\gamma + \gamma_0}{2})} \left [  -\Omega_{-}({\rho}_{cc}-{\rho}_{--})+ \Omega_{+}{\rho}_{-+} \right ]
 \label{eqnmcst}
\end{eqnarray}
respectively. 
Substituting Eqs. (\ref{eqnpcst}) and (\ref{eqnmcst}) in Eq. \ref{ro+-}, we obtain the steady-state value of 
\begin{eqnarray}
 {\rho}_{+-} = \frac{ \Omega_{+} \Omega_{-} [  w_{+}/\kappa_{+} - w_{-}/\kappa_{-}^{*}] }{4 (\Delta - i \gamma_0/2) -  [ \Omega_{-}^2 /\kappa_{+} -  \Omega_{+}^2/\kappa_{-}^{*} ] } 
 \label{eqnpmnew1}
\end{eqnarray}
where $\Delta = \delta_{+} - \delta_{-}$ is the two-photon detuning, $w_{+} = {\rho}_{++} - {\rho}_{cc}$, $w_{-} = {\rho}_{--} - {\rho}_{cc}$, $\kappa_{+} = \delta_{+} - i \frac{\gamma }{2}$ and $\kappa_{-} = \delta_{-} - i \frac{\gamma + \gamma_0 }{2}$ 
Substitution of Eq.(\ref{eqnpmnew1}) in Eq. (\ref{eqnpcst}) and  Eq. (\ref{eqnmcst}) leads to 
\begin{eqnarray}
 {\rho}_{+c} = \frac{1}{2 \kappa_{+}}  \Omega_{+} \left [  w_{+} +   \frac{  \Omega_{-}^2 [ w_{+}/\kappa_{+} - w_{-}/\kappa_{-}^{*}  ] }{4 (\Delta - i \gamma_0/2) -  [  \Omega_{-}^2 /\kappa_{+} - \Omega_{+}^2/\kappa_{-}^{*} ] } \right ]
 \label{eqnpcstnew1}
\end{eqnarray}
\begin{eqnarray}
 {\rho}_{-c} = \frac{1}{2 \kappa_{-}}   \Omega_{-} \left [  w_{-} +   \frac{  \Omega_{+}^2 [ w_{+}/\kappa_{+}^* - w_{-}/\kappa_{-}  ] }{4 (\Delta + i \gamma_0/2) -  [ \Omega_{-}^2 /\kappa_{+}^{*} -  \Omega_{+}^2/\kappa_{-} ] } \right ]
 \label{eqnmcstnew1}
\end{eqnarray}
By taking $w_{+}$ and $w_{-}$ outside the third bracket of Eqs. (\ref{eqnpcstnew1}) and (\ref{eqnmcstnew1}) respectively, both these equations  can be expressed in terms of the ratio $r = w_{-}/w_{+}$ of the two population inversions. Thus we have 
\begin{eqnarray}
 {\rho}_{+c} = \frac{w_{+}}{2 \kappa_{+}}   \Omega_{+} \left [  1 +   \frac{  \Omega_{-}^2 [ 1/\kappa_{+} - r/\kappa_{-}^{*}  ] }
 {4 (\Delta - i \gamma_0)  -  [  \Omega_{-}^2 /\kappa_{+} -  \Omega_{+}^2/\kappa_{-}^{*} ] } \right ]
 \label{eqnpcstnew}
\end{eqnarray}
\begin{eqnarray}
 {\rho}_{-c} = \frac{w_{-}}{2 \kappa_{-}}   \Omega_{-} \left [  1 +   \frac{  \Omega_{+}^2 [ 1 /r\kappa_{+}^* - 1/\kappa_{-}  ] }
 {4 (\Delta + i \gamma_0/2) -  [  \Omega_{-}^2 /\kappa_{+}^{*} -  \Omega_{+}^2/\kappa_{-} ] } \right ]
 \label{eqnmcstnew}
\end{eqnarray}
 Equation (\ref{eqncc}) provides 
\begin{eqnarray}
 \frac{ \Omega_{-} {\rm Im}[ {\rho}_{-c}] + \gamma_o {\rho}_{--}} {\Omega_{+} {\rm Im}[{\rho}_{+c}] - \gamma_0 {\rho}_{--}} =  \frac{\gamma_{-} }{ \gamma_{+}}
 \label{ratio}
\end{eqnarray}
Equations (\ref{eqnpcstnew}) and (\ref{eqnmcstnew}) show that the left hand side Eq. (\ref{ratio}) will be a function of $r$ and ${\rho}_{--}$. The population inversions $w_{+}$ and $w_{-}$ can be evaluated by using the Eq. (\ref{ratio}), (\ref{eqncc}) and the normalization condition.
$\rho_{cc} + \rho_{++} + \rho_{--} = 1$. 
\bibliographystyle{ieeetr}
\bibliography{NegativeMaterials,NIM2}

\end{document}